\documentclass[12pt]{article}
\usepackage{color} %
\usepackage{a41} %
\usepackage{float} %
\usepackage{lscape} %
\usepackage{epsfig} %
\usepackage{amssymb} %
\usepackage{amsmath} %
\usepackage{verbatim} %
\usepackage{cite} %
\newcommand{\N}{\nonumber}

\newcommand{\Li}{\mbox{Li}}

\newcommand{\Mvec}{\mbox{\rm\bf M}}

\newcommand{\beq}{\begin{equation}}
\newcommand{\eeq}{\end{equation}}
\newcommand{\bea}{\begin{eqnarray}}
\newcommand{\eea}{\end{eqnarray}}

\newcommand{\GeV}{${\rm GeV}$}

\graphicspath{/usr1/hboett/plots/qcd_pol/}
\graphicspath{./}

\begin{document}
\begin{titlepage}

\begin{flushleft}
DESY 11--032  
\\ 
DO--TH 11/06  \\
SFB-CPP/11--19\\
LPN 11--19 \\
April 2011 \\
\end{flushleft}

\vspace{1.5cm}
\noindent
\begin{center}
{\LARGE\bf \boldmath $O(\alpha_s)$ Heavy Flavor Corrections to} 

\vspace{2mm}
{\LARGE\bf \boldmath Charged Current Deep-Inelastic Scattering}

\vspace{2mm}
{\LARGE\bf \boldmath in Mellin Space
}
\end{center}
\begin{center}

\vspace{2cm}

{\large J. Bl\"umlein, A. Hasselhuhn, P. Kovacikova, S. Moch}

\vspace{1.5cm}
{\it Deutsches Elektronen--Synchrotron, DESY,\\
Platanenallee 6, D--15738 Zeuthen, Germany}

\vspace{2.5cm}
\end{center}

\begin{abstract}
\noindent
We provide a fast and precise Mellin-space implementation of the $O(\alpha_s)$ heavy
flavor Wilson coefficients for charged current deep inelastic scattering processes.
They are of importance for the extraction of the strange quark distribution in 
neutrino-nucleon scattering and the QCD analyses of the HERA charged current data.
Errors in the literature are corrected. We also discuss a series of more general parton 
parameterizations in Mellin space. 
\end{abstract}
\end{titlepage}

\vfill
\newpage
\sloppy

\section{Introduction}

\vspace*{1mm}
\noindent
The scale evolution in Quantum Chromodynamics (QCD) can be performed very efficiently
in Mellin space. There the evolution equations of the twist--2 parton distributions
and the corresponding observables become ordinary differential equations, unlike in
momentum-fraction space, where integro-differential equations have to be solved 
numerically. In Mellin space, moreover, the evolution equations can be solved 
analytically, provided that the anomalous dimensions and Wilson coefficients can be
represented at complex values of the Mellin variable $N$, cf.~\cite{EVO}. 
The transformation mediating between both representations is
\begin{eqnarray}
\label{eq:MEL}
\Mvec\left[f(x)\right](N) = \int_0^1 dx~x^{N-1}~f(x)~,
\end{eqnarray}
where $f(x)$ may  be distribution valued.
The anomalous dimensions 
and massless Wilson coefficients to 3--loop order are functions of the nested harmonic 
sums \cite{HSUM1,HSUM2}, see~Refs.~\cite{3LOOP}. The nested harmonic sums 
$S_{\vec{a}}(N)$ are 
meromorphic functions in the complex plane with poles at the non positive integers and
in case that the first $k$ indices $a_i$ are equal to one, they diverge $\propto 
\ln^k(N)$
for large values of $N$, \cite{STRUCT5}. They obey recursion relations for arguments 
$N \rightarrow (N+1)$ in terms of harmonic sums of lower weight, which allow for shifts
parallel to the real axis. Moreover, analytic asymptotic representations have been 
derived,
\cite{STRUCT5,STRUCT6}, which are valid in the region of large values of $N, 
|\text{arg}(N)|  < \pi$. One may derive effective numerical representations using 
the {\tt MINIMAX}-method \cite{MINIMAX}, cf. \cite{ANCONT,ANCONT1}. In this way the 
QCD 
observables at the level of twist-2 are represented analytically at general scales 
$Q^2$, parameterizing the parton density functions by appropriate non-perturbative 
distributions at a starting scale $Q_0^2$. This also applies for a wide class of other 
processes, cf. \cite{MELLINSP}.   

In many precision analyses  heavy quark  contributions have to be 
considered, which are more difficult to represent in Mellin space. For deep-inelastic 
scattering the contributions up to the $O(\alpha_s^2)$ neutral current heavy flavor 
Wilson coefficients \cite{HEAV1}, which are available in semi-analytic form in momentum 
fraction space, were given in Mellin space in Ref.~\cite{AB}. The running mass effects
have been described in \cite{RUNMQ}.

In the present note
a fast and precise implementation of the $O(\alpha_s)$ charged current heavy-flavor
Wilson coefficients for deep-inelastic scattering is presented. In Section~2 we derive 
the corresponding Mellin-space representations for the charged current structure 
functions up to $O(\alpha_s)$. As a significant part of the data emerges for large values 
of $Q^2$, we also derive the representation in Section3 which is valid for $Q^2 \gg m_c^2$, with $Q^2 = 
-q^2$, $q$ being the four-momentum transfer, and $m_c$ the charm quark mass. We illustrate 
the precision of the Mellin-space implementation by comparing to the structure 
functions in momentum fraction space in Section 4 . Finally we remark on the Mellin space 
implementations of 
various $x$-space distributions used in different analyses in the literature in Section 5, which may 
be important to account for a higher flexibility in the choice of the non-perturbative 
input distributions. This will allow to perform analogous analyses in Mellin space in 
the future. 
\section{The Scattering Cross Section}

\vspace*{1mm}
\noindent
The charged current deep-inelastic scattering cross section for 
$\nu (\overline{\nu}) p \rightarrow l (\overline{l}) +X,~~~ 
l (\overline{l})  p \rightarrow \nu (\overline{\nu}) +X$ 
in case of charm-quark production is given by
\begin{eqnarray}
\label{eq:XS}
\frac{d^2 \sigma}{dx dy} &=& \frac{G_F^2 S}{4 \pi}\frac{M_W^4}{(M_W^2 + Q^2)^2}
\Biggl\{y^2 2x\mathbb{F}_1^c(x,Q^2,m_c^2) + 2 \left(1 - y - y^2 \frac{M^2 x^2 
}{Q^2}\right) 
\mathbb{F}_2^c(x,Q^2,m_c^2)
\nonumber
\end{eqnarray}
\begin{eqnarray}
&& \pm  \left[1-(1-y)^2\right] x\mathbb{F}_3^c(x,Q^2,m_c^2) \Biggr\}. 
\end{eqnarray}
Here $x = Q^2/yS, y = q.P/l.P$ are the Bjorken variables, 
$l$ and $P$ are the lepton and nucleon momentum, $S = 2l.P$, $G_F$ is the Fermi 
constant,
$M_W$ the mass of the $W$-boson, $M$ the nucleon mass, and $\mathbb{F}_i(x,Q^2)$ are 
the structure functions. The $\pm$ signs in 
(\ref{eq:XS}) refer to incoming neutrinos (anti-neutrinos) or charged leptons 
(anti-leptons), respectively. Note that the scattering cross section does depend on 
the 
masses of the initial state particles, cf. \cite{HECTOR}. Due to the inclusive kinematics 
the dependence on $m_c$ is only implicit.

In the twist-2 approximation, invoking the parton model, one may decompose the 
nucleon 
wave function into individual partons and consider the excitation of single charm quarks
in the transitions
\begin{eqnarray}
{s}' = 
s~|V_{cs}|^2 + 
d~|V_{cd}|^2  \rightarrow  c,~~~~~~~~~~
\bar{s}' = 
\bar{s}~|V_{cs}|^2 + 
\bar{d}~|V_{cd}|^2  \rightarrow  \bar{c}~,
\end{eqnarray}
with $V_{ij}$ the CKM matrix elements \cite{CKM}, and $s$ and $d$ the strange and 
down quark 
parton distributions. Here $\bar{d}$ and $\bar{s}$  denote the corresponding anti-quark 
distributions. For this transition 
the Bjorken variable $x$ and the momentum fraction of the struck parton $\xi$ are 
related by
\begin{eqnarray}
 x = \xi \lambda \leq \lambda~.
\end{eqnarray}
with $\lambda = {Q^2}/(Q^2 +m_c^2)$. This phenomenon is called slow rescaling 
\cite{BROCK}. The heavy quark structure functions and parton densities can be expressed 
as a function of $\xi \in [0,1]$, preserving Mellin symmetry.

The $O(\alpha_s)$ charged current heavy-flavor Wilson coefficients were calculated in
\cite{GOTTSCH} and corrected in \cite{DORTM1} later. We will follow Ref.~\cite{DORTM1} 
and work in the fixed flavor number scheme (FFNS).~\footnote{The use of a variable 
flavor number scheme needs care, since the scales at which one massive flavor can 
be dealt with as massless is process dependent. The corresponding scales are 
usually not $\mu^2 \sim m^2$, cf. Ref.~\cite{BN}.}
We define
\begin{eqnarray}
  {\mathcal F}_1^c=\mathbb{F}_1^c,\quad {\mathcal 
F}_2^c=\mathbb{F}_2^c/2\xi,\quad{\mathcal 
F}_3^c=\mathbb{F}_3^c/2 \end{eqnarray} 
To $O(\alpha_s)$ one obtains, after a Mellin transformation (\ref{eq:MEL}) over 
$\xi$, 
\begin{eqnarray} {\mathcal F}_i^c(N,Q^2) &=& 
{s'}(N,\mu^2)+ a_s 
  \Biggl[H_i^{(1),q}\left(N,\frac{m^2}{\mu^2},\frac{Q^2}{\mu^2}\right)
  {s'}(N,\mu^2) 
+ H_i^{(1),g}\left(N,\frac{m^2}{\mu^2},\frac{Q^2}{\mu^2}\right)
  g(N,\mu^2) \Biggr]~,
\nonumber\\
\end{eqnarray}
for $W^+$ exchange. In case of $W^-$ exchange $\bar{s}'$ replaces $s'$. Here
$a_s = \alpha_s/(4\pi)$ denotes the strong coupling constant and $g$ the gluon 
distribution.
The massive Wilson coefficients for charged current deep-inelastic scattering 
in Mellin space are given by\footnote{Throughout the present paper we consider
the scale derivative in the renormalization group operator  as $\partial/\partial 
\ln(\mu^2)$.}
\begin{eqnarray}
H_i^q\left(N,\frac{m^2}{\mu^2},\frac{Q^2}{\mu^2}\right)
&=& 1 + \sum_{k=1}^\infty a_s^k
H_i^{(k),q}\left(N,\frac{m^2}{\mu^2},\frac{Q^2}{\mu^2}\right)
\\
H_i^g\left(N,\frac{m^2}{\mu^2},\frac{Q^2}{\mu^2}\right)
&=& \sum_{k=1}^\infty a_s^k
H_i^{(k),g}\left(N,\frac{m^2}{\mu^2},\frac{Q^2}{\mu^2}\right)
\end{eqnarray}
with 
\begin{eqnarray}
H_{i}^{(1),q}\left(N,\frac{m^2}{\mu^2},\frac{Q^2}{\mu^2}\right)
&=& \frac{1}{2} P_{qq}^{(0)}(N) \ln \left(\frac{Q^2 + m_c^2}{\mu^2}\right) + 
h_i^q(\lambda,N),~~i = 1,2,3,
\end{eqnarray}
\begin{eqnarray}
H_{1,2}^{(1),g}\left(N,\frac{m^2}{\mu^2},\frac{Q^2}{\mu^2}\right)
&= & \frac{1}{4}
P_{qg}^{(0)}(N)
\ln\left(\frac{Q^2+m_c^2}{\mu^2}\right) + \frac{1}{4} \tilde{P}(\lambda,N)
+ h_{1,2}^g(\lambda,N)
\\
\label{eq:H3M}
H_{3}^{(1),g}\left(N,\frac{m^2}{\mu^2},\frac{Q^2}{\mu^2}\right)
&= & \frac{1}{4}
P_{qg}^{(0)}(N)
\ln\left(\frac{Q^2+m_c^2}{\mu^2}\right) - \frac{1}{4}
\tilde{P}(\lambda,N)
+ h_{3}^g(\lambda,N)~.
\end{eqnarray}
Here the leading order splitting functions are
\begin{eqnarray}
{P}_{qq}^{(0)}(N) &=& 4 C_F~\left[\frac{3}{2} + \frac{1}{N(N+1)} - 2 S_1(N)\right]
\\
{P}_{qg}^{(0)}(N) &=& 8 T_F~\frac{N^2+N+2}{N(N+1)(N+2)}~,
\end{eqnarray}
with $T_F = 1/2, C_F = (N_c^2-1)/(2 N_c)$ for $SU(N_c)$ and $N_c = 3$ for QCD.
The function $\tilde{P}(\lambda,N)$ is given by
\begin{eqnarray}
\tilde{P}(\lambda,N) &=&
\Mvec\left[P_{qg}^{(0)}(z)\ln\frac{1-\lambda z}{(1-\lambda)z}\right](N) \nonumber\\
&=&
8 T_F \Biggl\{\frac{\left[\lambda^2 \left(N^2+3 N+2 \right) -2 \lambda \left(N^2+2
N\right)+2 N^2+2 N\right] }{\lambda N (N+1) (N+2)} \left[ F_2(\lambda,N)
- \frac{1}{\lambda} \ln(1-\lambda)\right]
\N\\
& &+\frac{\lambda \left(N^2+3  N+2 \right)-2 N^2}{\lambda N^2 (N+1) 
(N+2)}\Biggr\}~.
\end{eqnarray}
In the limit $\lambda \rightarrow 1$ it takes the form
\begin{eqnarray}
\tilde{P}(1,N) &=& 8 T_F\left\{
\frac{N^2 + N + 2}{N(N+1)(N+2)} \left[ \ln \left(\frac{Q^2}{m^2}\right) - 
S_1(N)\right] 
+ \frac{-N^2 + 3 N + 2}{N^2(N+1)(N+2)}\right\}~.
\end{eqnarray}

The functions ${h}_{i}^{q(g)}(\lambda,N)$ read~:
\begin{eqnarray}
\label{heq1}
{h}_1^{q}(\lambda,N) 
&=& C_F \Biggl[
4 F_1(\lambda,N)
- \frac{\left[\lambda^2 \left( N^3+ N+2 \right)
               -\lambda \left(N^3+3N^2+ 4 N\right) 
               +2 N^2+2 N\right] }{\lambda N (N+1)} F_2(\lambda,N)
\N\\
& &
+ \frac{4 \left(N^2+N-1\right) }{N (N+1)} S_1(N)
+ 4S_1^2(N)
- \frac{(1-\lambda) \left[\lambda \left(N^2+3  N+4 \right)-2 N-2\right] }
  {\lambda^2 (N+1)} \ln (1-\lambda)
\N\\
& &-\frac{\left[\lambda( 9 N^3+8  N^2-5  N-2) -2 N^2\right]}{\lambda N^2 (N+1)}
\Biggr]
\\
{h}_2^{q}(\lambda,N)
&=& C_F\Biggl[
  4  F_1(\lambda,N) 
+ \frac{\left[4 \lambda^2 \left( N^2+ N\right) -\lambda \left(N^3+2  N^2+3  N+2 
  \right)+N^3-N^2\right] }{N (N+1)} F_2(\lambda,N)
\N\\
& &
+ \frac{4\left(N^2+N-1\right) }{N (N+1)} S_1(N) 
+ 4S_1^2(N)
+ \frac{(1 - \lambda) \left[4 \lambda (N+1) -N^2 
+N \right] }{\lambda (N+1)} \ln (1-\lambda)
\N\\
& &
+\frac{\left[4  \lambda (
 N^2 + N) -9 N^3-6 N^2+N+2\right]}{N^2 (N+1)} \Biggr] 
\end{eqnarray}
\begin{eqnarray}
{h}_3^q(\lambda,N)
&=& C_F \Biggl[
  4 F_1(\lambda,N)
- \frac{ (N-1) \left[\lambda (N^2-N-2)-N^2\right] }{ N (N+1)} F_2(\lambda,N)
+ \frac{4 \left(N^2+N-1\right) }{ N (N+1)} S_1(N)
\N\\
& &
+ 4 S_1^2(N)
- 
\frac{ (1-\lambda) \left(N^2+N+2\right) }{ \lambda (N+1)}
\ln(1-\lambda)
- \frac{ (3 N+2) \left(3 N^2-1\right)}{ N^2 (N+1)}
\Biggr]
\\
{h}_1^g(\lambda,N)
&=&
- 4 T_F \Biggl[
\frac{\left[\lambda^2 (2 N^3+3 N^2+7  N+2 ) -\lambda(2 
N^3+4  N^2+8 N)+2 N^2+2 N\right] }{2 \lambda N (N+1) (N+2)} F_2(\lambda,N)
\N\\
& &
+ \frac{\left(N^2+N+2\right) }{N (N+1) (N+2)} S_1(N)
+ 
\frac{(1-\lambda) 
\left[\lambda (N^2+N+3)-N-1\right] }{\lambda^2 (N+1) (N+2)}
\ln(1-\lambda)
\N\\
& &
+ \frac{\lambda(2 N^3 + N^2+ N-2)-2 N^2}{2 \lambda N^2 
(N+1) (N+2)} \Biggr]
\\
{h}_2^g(\lambda,N)
&=& 
- 4 T_F \Biggl\{\Biggl[\frac{\left[\lambda^3 (2 N^3-6  N^2+4  
N) + \lambda^2 (-2  N^3+7  N^2- N+2) - \lambda( 2  N^2
+ 4 N)\right] }{2 \lambda N (N+1) (N+2)} 
\N\\ & & + \frac{1}{\lambda(N+2)}\Biggr] 
F_2(\lambda,N)
+
\frac{(1-\lambda) 
\left[\lambda^2 (N^2-3 N+2)+\lambda-N-1\right] }
{\lambda^2 (N+1) (N+2)}\ln(1-\lambda)
\N\\ 
& &
+ \frac{\left(N^2+N+2\right) }{N (N+1) (N+2)} S_1(N)
+\frac{\lambda^2( 2 N^3-6 N^2+4  N)-\lambda (N^2+3 N)-2 
\lambda-2 N^2}{2 \lambda N^2 (N+1) (N+2)} \Biggr\}
\nonumber\\
\\
\label{heq6}
{h}_3^g(\lambda,N)
&=&
- 4 T_F \Biggl[\frac{\left(\lambda^2 (N^2- N+2) + \lambda (2 N^2+4  N)-2 N^2
  -2 N\right) }{2 \lambda N (N+1) (N+2)} F_2(\lambda,N)
\N
\end{eqnarray}
\begin{eqnarray}
& &
+ \frac{\left(N^2+N+2\right) }{N (N+1) (N+2)} S_1(N)
- \ln (1-\lambda)\frac{(1- \lambda) (\lambda-N-1) }{\lambda^2 (N+1) (N+2)}
\nonumber\\
&&
- \frac{\lambda (N^2+3  N+2)-2 N^2}{2 \lambda N^2 (N+1) (N+2)}
\Biggr]~.
\end{eqnarray}
Here $S_k(N)$ are the single harmonic sums \cite{HSUM1,HSUM2}
\begin{eqnarray}
  S_k(N) = \frac{(-1)^{k+1}}{(k-1)!} 
\psi^{(k-1)}(N+1) +\zeta_k, k \in \mathbb{N} \backslash \{0\}
\end{eqnarray}
and $F_{1,2}(\lambda,N)$ denote the functions 
\begin{eqnarray}
\label{eqf1}
  F_1(\lambda,N) &=& \int_0^1 dx\; \frac{x^N-1}{x-1}\ln(1-\lambda x) \\
\label{eqf2}
  F_2(\lambda,N) &=& \int_0^1 dx\; \frac{x^N-1}{1-\lambda x}~.
\end{eqnarray}

We refer to $H_2^{(1),g}$ for $Q^2 = \mu^2$ also in the form
\begin{eqnarray}
H_2^{(1),g}(\lambda,N) &=&
4 T_F 
\Biggl \{
\frac{4-2N(N-3)-N(N^2+N+2)\lbrace 
2S_1(N)+\ln[\lambda(1-\lambda)]\rbrace}{2N^2(N+1)(N+2)}
\N
\end{eqnarray}
\begin{eqnarray}
& &
+\frac{8-18(1-\lambda)+12(1-\lambda)^2}{(N+1)(N+2)}
+\frac{(1-\lambda) _2F_1(1,N,N+1;\lambda)}{N}-\frac{1}{N}
\N\\& &
+6\lambda(1-\lambda)\Biggl
[\frac{_2F_1(1,N+1,N+2;\lambda)}{(N+1)^2}
-2 \frac{_2F_1(1,N+1,N+2;\lambda)-1}{(N+1)(N+2)}\Biggr]\Biggr\}
\nonumber\\
\end{eqnarray}
to correct Eq.~(70) in Ref.~\cite{Ball:2011mu}.
Here, $_2F_1$ denotes the Gau\ss{}' hypergeometric function
\begin{eqnarray}
\label{eq:2f1}
_2F_1(\eta,N+\alpha+1, N + \alpha + \beta 
+ 2; \xi) = \frac{1}{B(N + \alpha + 1, \beta +1)}
\int_0^1~dz z^{N+\alpha} (1-z)^\beta (1 - \xi z)^{-\eta}. 
\nonumber\\
\end{eqnarray}
\restylefloat{table}
\begin{table}[H]  
\small
\begin{center}
\renewcommand{\arraystretch}{1.5}
\begin{tabular}{|r|l|l|l|}
\hline
$k \backslash \lambda$ & $0.1$ & $0.5$ & $0.9$ \\
\hline
 0  &   +9.999999999999999999
    &   +1.999999999999954543 
    &   +1.111110872017648708 
\\
 1  &   +0.821423460E-22 
    &   +0.1975123084486687E-10 
    &   +0.000098847800695649 
\\
 2  &   $-$0.050000000000000000 
    &   $-0.250000001423445310$ 
    &   $-0.456776649795426910$  
\\
 3  &   $-0.001666666666666666$
    &   $-0.041666626227537189$
    &   +0.048001787041481484
\\
 4  &   $-0.000083333333333336$
    &   $-0.010417267014473893$
    &   $-2.639926313632337045$
\\
 5  &   $-$0.499999999997477753E-05
    &   $-$0.003119649969153064 
    &   +21.74650303005800508  
\\
 6  &   $-$0.333333333484935779E-06
    &   $-$0.001072643292472726
    &   $-$119.2024353694977123  
\\
 7  &   $-$0.238095231857886535E-07 
    &   $-$0.000249816388951957
    &   +442.8630812379568060
\\
 8  &   $-$0.178571609458723466E-08 
    &   $-$0.000477192766818736 
    &   $-$1147.492168730891145
\\
 9  &   $-$0.138885135219891312E-09 
    &   +0.000607438245655042 
    &   +2095.900562359927839  
\\
10  &   $-$0.111166985797600287E-10 
    &   $-$0.000939620170907808 
    &   $-$2688.830619105423874   
\\
11  &   $-$0.903191209860331475E-12 
    &   +0.000876266420390668 
    &   +2371.811788035594662
\\
12  &   $-$0.800460593905461726E-13 
    &   $-$0.000571683728308967 
    &   $-$1370.383159208738358 
\\
13  &   $-$0.439528295242656941E-14 
    &   +0.000219008887781703 
    &   +467.1985539745224063    
\\
14  &   $-$0.107994848505839019E-14 
    &   $-$0.000041032031472769
    &   $-$71.30766062862566906
\\
\hline
\end{tabular}
\renewcommand{\arraystretch}{1.5}
\caption[]{\sf The expansion coefficients $a_k$, Eq.~(\ref{eq:MI2}) for special values of $\lambda$.}
\end{center}
\end{table}

Many of the contributing functions have known Mellin transforms given before in \cite{HSUM1,HSUM2}.
Their analytic continuations to complex values of $N$ are known, 
cf.~\cite{STRUCT5,STRUCT6,ANCONT,ANCONT1}.
The new functions (\ref{eqf1},\ref{eqf2}) are solely related to integrals of the kind 
(\ref{eq:2f1}) for $N \in \mathbb{N}$. They obey the following recursion relations 
\begin{eqnarray}
\label{eq:di1}
  F_1(\lambda,N+1) &=& F_1(\lambda,N) + \frac{\lambda}{N+1} F_2(\lambda,N+1)  \\
\label{eq:di2}
  F_2(\lambda,N+1) &=& \frac{1}{\lambda} \left[F_2(\lambda,N) - \frac{1}{N+1} - \frac{1-\lambda}{\lambda} 
  \ln(1-\lambda)\right]~.
\end{eqnarray}
The singularity structure of $F_{1,2}(\lambda,N)$ for $N \in \mathbb{C}$ can be seen 
using the representation
\begin{eqnarray}
F_1(\lambda,N) &=& - \sum_{l=1}^\infty \frac{\lambda^l}{l} \left[\psi(l+N+1) - 
\psi(l+1)\right] \\
F_2(\lambda,N) &=&  - N \sum_{l=0}^\infty \frac{\lambda^l}{(N+l+1)(l+1)}~,
\end{eqnarray}
where $\psi^{(k)}(z), k \geq 0$ denotes the polygamma function.
Both functions possess poles in $N$ at negative integers.

For $F_1(\lambda,N)$ one may derive a sufficiently precise representation using the 
{\tt MINIMAX}-method. The function $F_1$ can be written as
\begin{eqnarray}
\label{eq:MI1}
    F_1(\lambda,N) &=& -\left\{
     \frac{1}{\lambda}-N \mathcal{E}(\lambda,1)
     +\sum_{l=0}^{14} a_l~\left\{N -1 - l[S_1(N+l-1)-S_1(l)]\right\}\right\}~, \\
  \mathcal{E}(\lambda,x) &\equiv&   
    \frac{1}{\lambda}(1-\lambda x)(1-\ln(1-\lambda x))~.
\end{eqnarray}
The function $\mathcal{E}$ under the integral is 
approximated by the adaptive polynomial, cf. Table~1,
\begin{eqnarray}
\label{eq:MI2}
   \mathcal{E}(\lambda,x) = \sum_{k=0}^{14} a_k(\lambda)~x^k~.
\end{eqnarray}
Knowing the difference equations (\ref{eq:di1},\ref{eq:di2}) one may shift $F_{1,2}$ parallel to the real axis.
Usually one attempts to shift towards the asymptotic region and applies an analytic representation there.
However, as sometimes recommended, one may also turn the view \cite{COP}, and 
rather consider 
$F_{1,2}(\lambda,N)$ inside the unit circle, to obtain an even more beautiful result~:  
\begin{eqnarray}
\label{eq:F1U}
F_1(\lambda,N) &=& - \frac{1}{2} \left[S_1^2(N) + S_2(N)\right] \nonumber
\\
&& 
- \sum_{k=1}^\infty (-N)^k \left\{ \Li_{k+2}(\lambda) - \zeta_{k+2} + 
H_{1,k+1}(\lambda) - H_{1,k+1}(1)\right\}\\
\label{eq:F2U}
F_2(\lambda,N) &=& \frac{1}{\lambda} \left\{\sum_{k=1}^\infty (-N)^k 
\left[\Li_{k+1}(\lambda) - 1 \right] 
- \frac{N}{N+1}\right\}, 
\end{eqnarray}
for $|N| < 1$. Here, $\Li_k(x)$ denotes the classical polylogarithm \cite{LEWIN}, 
$H_{\vec{a}}(x)$ a harmonic polylogarithm \cite{VR}, 
\begin{eqnarray}
H_{1,k}(x) := \int_0^x~dz~\frac{\Li_k(z) - \zeta_k}{1-z}~,
\end{eqnarray}
and $\zeta_k$ Riemann's 
$\zeta$-function at an integer argument.
In the limit $\lambda \rightarrow 1$ one obtains
\begin{eqnarray}
  F_1(1,N) &=& - \frac{1}{2}\left[S_1^2(N) + S_2(N)\right] \\
  F_2(1,N) &=& - S_1(N)~.
\end{eqnarray}

The serial representations (\ref{eq:F1U}, \ref{eq:F2U}) are quickly converging and 
deliver even more precise results than using (\ref{eq:MI1}, \ref{eq:MI2}). The contour 
integral is performed along the line $(-\infty,-a),(c,-a),(c,a),(-\infty,a)$, with 
$a=0.5$. The choice of the parameter $c$ depends also on the rightmost singularity of 
the non-perturbative distribution $f(x)$ and was chosen with $c = 1.5$ in the 
figures given below. The inverse Mellin transform is obtained by
\begin{eqnarray}
    f(x) = \frac{1}{\pi}\left\{ - \int_{-\infty}^{c} {\sf Im}\left[x^{-s-ia} 
\Mvec[f](s+ia)\right] ds
+\int_{0}^{a} {\sf Im}\left[i x^{-i s-c} \Mvec[f](is+c)\right]ds \right\}~.
\end{eqnarray}
We extended the integral to 1000 units parallel to the real axis. The recursion 
relations (\ref{eq:di1}, \ref{eq:di2}) together with the series (\ref{eq:F1U}, 
\ref{eq:F2U}) 
allow to compute $F_{1,2}(\lambda, N)$ along the integration contour. Here usually
the first 30 terms in the infinite sums (\ref{eq:F1U}, \ref{eq:F2U}) provide 
sufficient accuracy. 
\section{The Massive Wilson Coefficients in the Asymptotic Region}

\vspace*{1mm}
\noindent
In many applications the value of $\lambda$ becomes close to one. Within this
kinematic region $Q^2 \gg m_c^2$ the Wilson coefficients take a simpler form, and
the functions needed to express them are only harmonic sums. This has been shown
up to 3-loop order in case of neutral current deep-inelastic scattering in \cite{HQ}.
At $O(\alpha_s)$ one obtains the 
following relations~:
\begin{eqnarray}
\label{eqH1}
H_i^{q,(1)}\left(N, \frac{Q^2}{m^2},\frac{m^2}{\mu^2}\right) 
&=& 
C_{i,q}^{(1)}\left(N,\frac{Q^2}{\mu^2}\right)~,~~~i = 1,2,3
\\
\label{eqH2}
H_i^{g,(1)}\left(N, \frac{Q^2}{m^2},\frac{m^2}{\mu^2}\right) 
&=& \frac{1}{2} A_{Qg}^{(1)}\left(N, \frac{m^2}{\mu^2}\right) + 
C_{i,g}^{(1)}\left(N,\frac{Q^2}{\mu^2}\right)~,~~~i = 1,2 
\\
H_3^{g,(1)}\left(N, \frac{Q^2}{m^2},\frac{m^2}{\mu^2}\right) 
&=& - \frac{1}{2} A_{Qg}^{(1)}\left(N, \frac{m^2}{\mu^2}\right)~.  
\end{eqnarray}
Here, $A_{lm}^{(1)}$ denote the 1--loop massive operator matrix elements derived in the neutral 
current case and $C_{i,q(g)}^{(1)}$ are the massless 1-loop Wilson coefficients, 
\cite{FP1}.
Eqs.~(\ref{eqH1}, \ref{eqH2}) are derived by expanding (\ref{heq1}--\ref{heq6}) for 
$\lambda \rightarrow 1$. At $O(\alpha_s)$ there is no pure-singlet contribution.
It turns out, that $A_{qq}^{NS,(1)}$ vanishes as expected, because closed massive 
fermion loop contributions can occur at $O(\alpha_s^2)$ earliest. In the charged 
current case the interaction transmutes the massless $s'$-quark into the massive 
$c$-quark. The 
massless quark-loop contributions to $A_{Qg}^{(1)}$, occurs with a combinatorial 
factor $1/2$ compared to the neutral current case. These terms vanish, because the 
corresponding diagrams are scaleless. Because of this no quarkonic operator matrix 
element contributes to (\ref{eqH1}). Eqs.~(\ref{eqH1}, \ref{eqH2}) agree with 
Ref.~\cite{BUZA2}.
 
The massless Wilson coefficients obey
\begin{eqnarray}
C_{i,j}\left(N,\frac{Q^2}{\mu^2}\right) = \delta_{jq} + \sum_{k=1}^\infty a_s^k 
C_{i,j}^{(k)}\left(N,\frac{Q^2}{\mu^2}\right),
\end{eqnarray}
where 
\begin{eqnarray}
C_{i,q}^{(1)}\left(N,\frac{Q^2}{\mu^2}\right) &=& \frac{1}{2} P_{qq}^{(0)}(N) 
\ln\left(\frac{Q^2}{\mu^2}\right) + c_{i,q}^{(1)}(N) \\
C_{i,g}^{(1)}\left(N,\frac{Q^2}{\mu^2}\right) &=& \frac{1}{2} P_{qg}^{(0)}(N) 
\ln\left(\frac{Q^2}{\mu^2}\right)
 + c_{i,g}^{(1)}(N) 
\end{eqnarray}
\begin{center}
\begin{figure}[t] 
\epsfig{figure=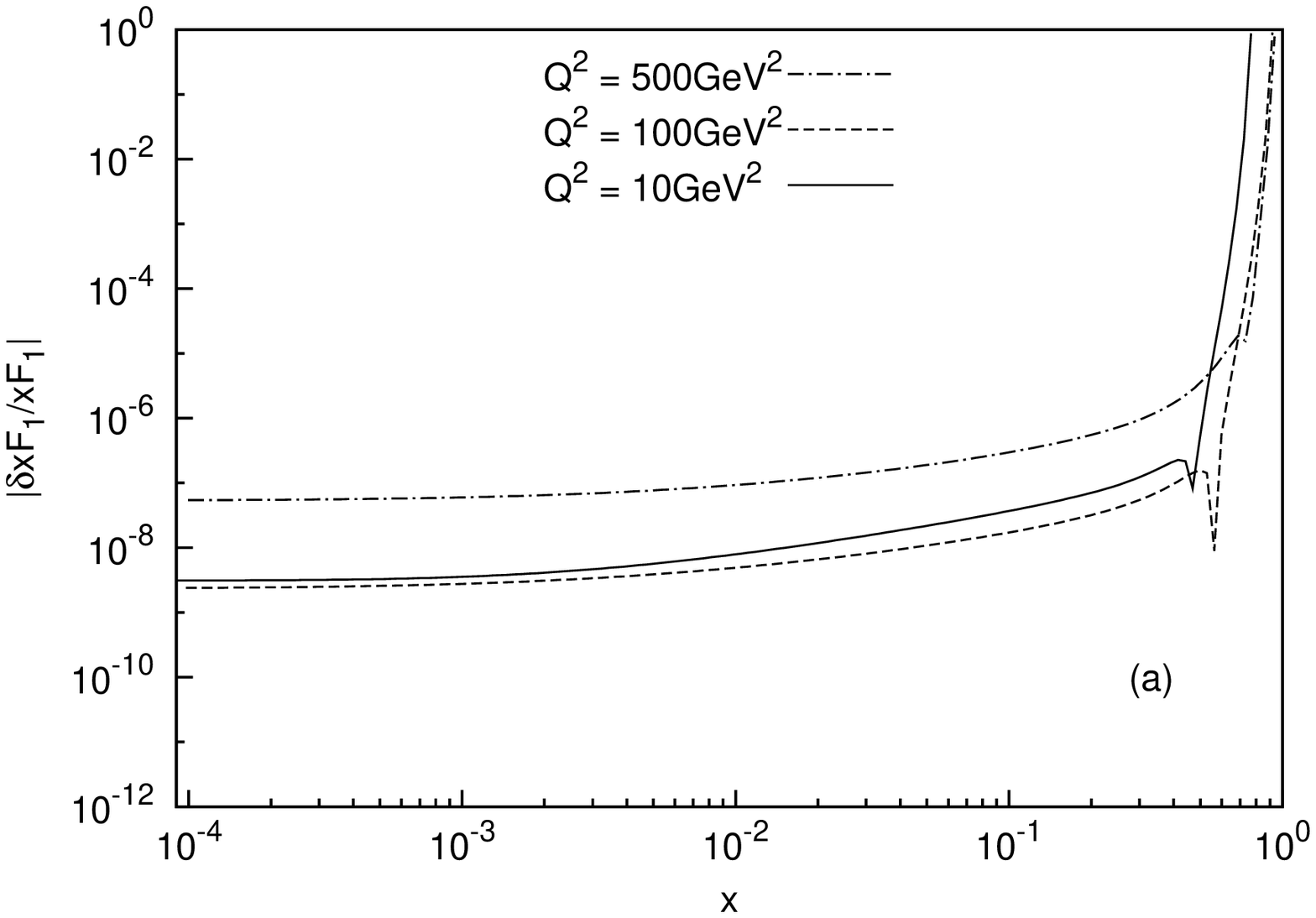,angle=-90,width=0.48\linewidth} \hspace*{2mm}
\epsfig{figure=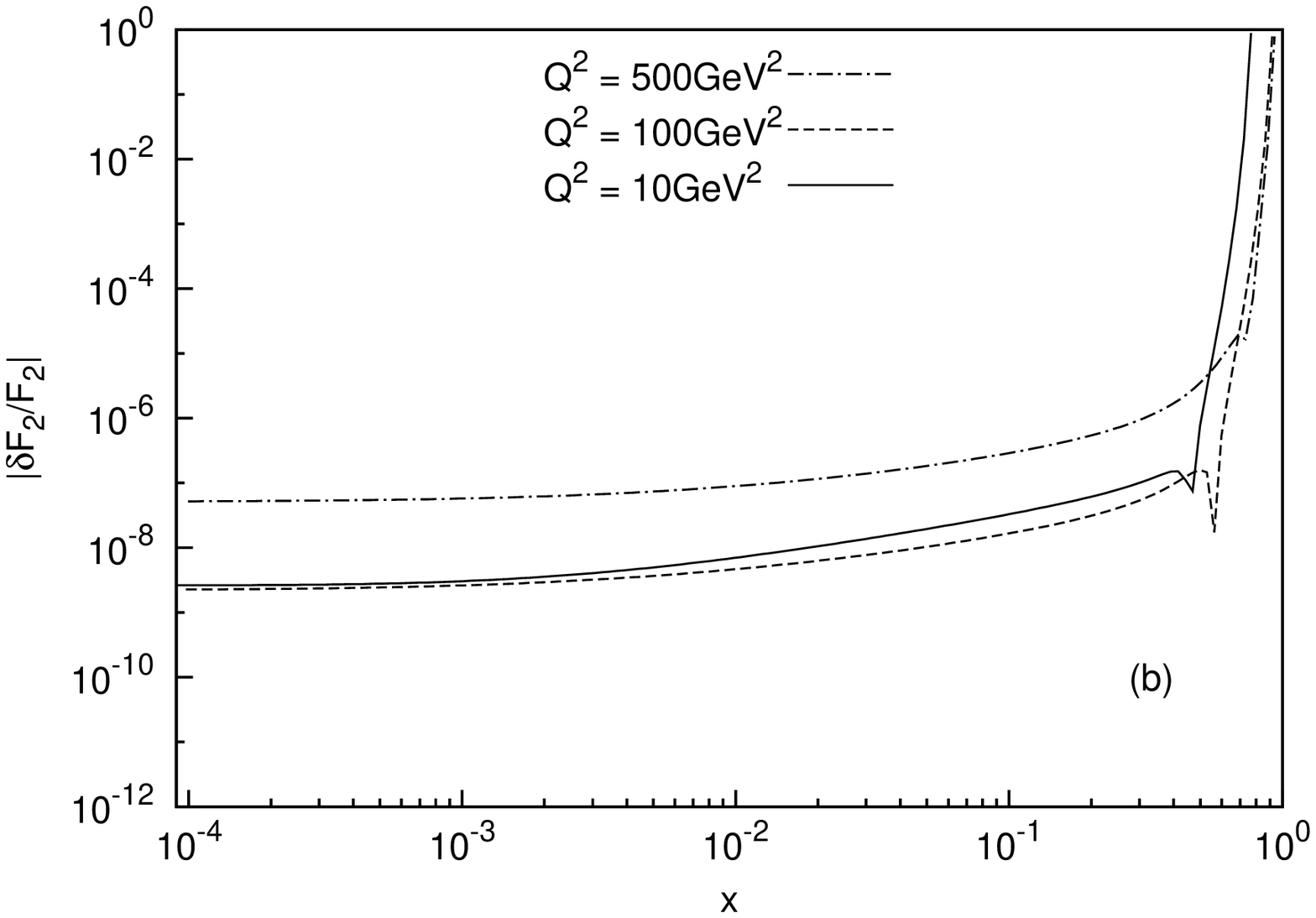,angle=-90,width=0.48\linewidth} 
\\
\begin{center}
\epsfig{figure=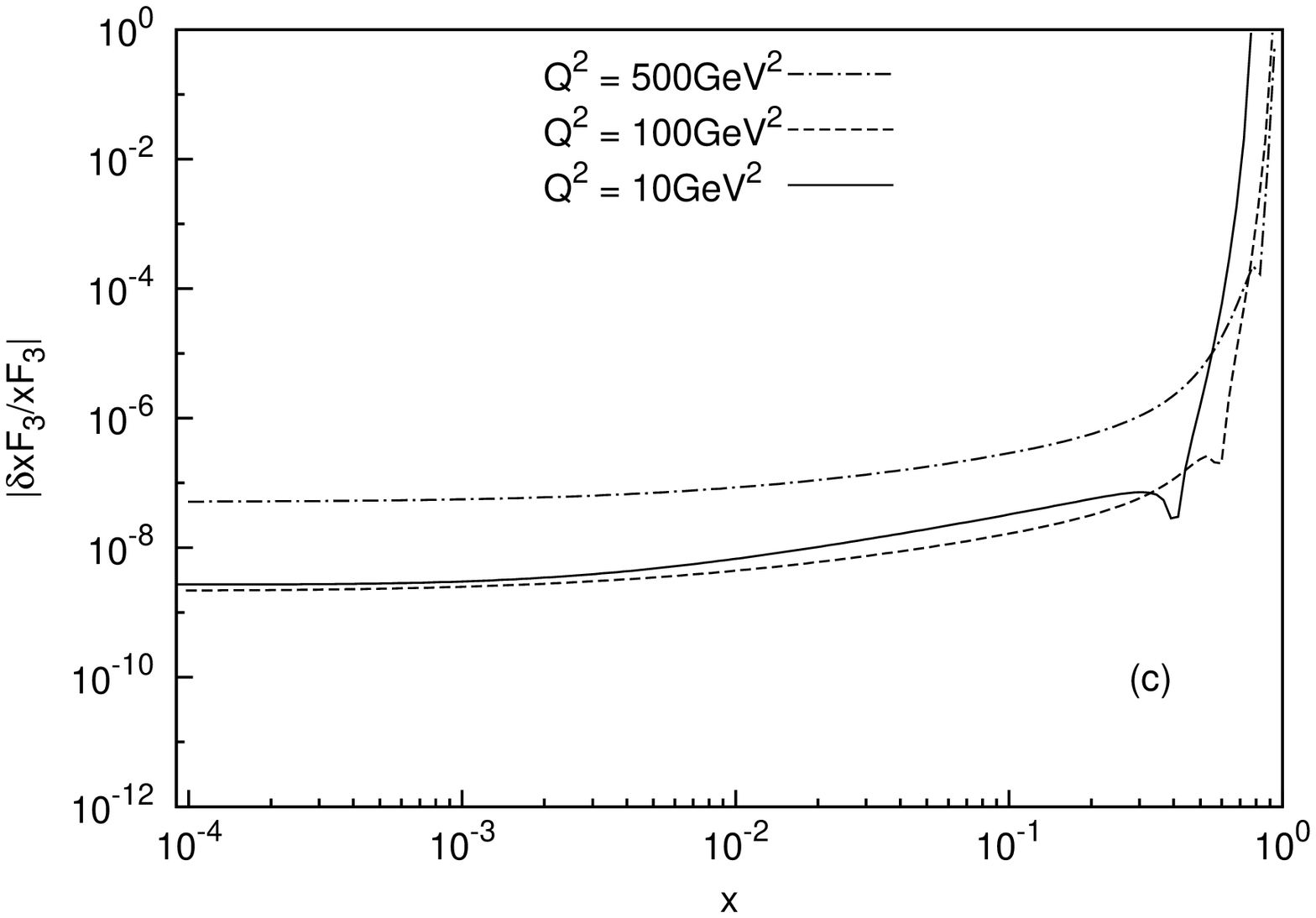,angle=-90,width=0.48\linewidth} 
\end{center}
\caption[]{
\label{FIG:ACC1}
\sf 
Relative accuracy of the charged current structure functions $\mathbb{F}_i$ 
to $O(\alpha_s)$ at $Q^2 = 10, 100, 500~\GeV^2$, comparing the implementation in 
Mellin- 
and $z$-space. Here both for $F_{1,2}(\lambda,N)$ the analytic representation 
(\ref{eq:F1U}, \ref{eq:F2U}) and their recurrences were used.}
\end{figure}
\end{center}
and \cite{FP}
\begin{eqnarray}
{c}_{1,q}^{(1)}(N) &=& {c}_{2,q}^{(1)}(N) - 4 C_F\frac{1}{N+1} 
\\
{c}_{2,q}^{(1)}(N) &=& C_F \Biggl[
\frac{\left(3N^2+3 N-2\right)}{N (N+1)} S_1(N) + 2 S_1^2(N)
- 2S_2(N)
-\frac{\left(9 N^3+2 N^2-5 N-2\right)}{N^2 (N+1)} \Biggr]
\nonumber\\
\\
{c}_{3,q}^{(1)}(N) &=& {c}_{2,q}^{(1)}(N) - C_F \frac{2(2N+1)}{N(N+1)} 
\\
{c}_{1,g}^{(1)}(N) &=& {c}_{2,g}^{(1)}(N) - T_F \frac{16}{(N+1) (N+2)}
\\
{c}_{2,g}^{(1)}(N)&=& T_F\Biggl[
-\frac{4 \left(N^2+N+2\right)}{N (N+1) (N+2)} S_1(N) -\frac{4(N^3-4 N^2-N-2)}{N^2 
(N+1) (N+2)}\Biggr]
\\
{c}_{3,g}^{(1)}(N)&=& 0~.
\end{eqnarray}
We note that the global sign for Eq. (A1.17) of \cite{BUZA2} has to be reversed.
The gluonic contribution in charged current heavy quark production has been 
calculated before using  two finite quark masses 
for $e^-p$ scattering in Refs.~\cite{WTB1,WTB2,WTB3}
and both for $e^-N$ and 
$e^+N$ scattering in \cite{WTB3}. For calculations in case of neutrino-nucleon 
scattering see \cite{nuN}. We repeated the calculation of the gluonic contribution for a 
massless $s$-quark in the $\overline{\rm MS}$-scheme.
\restylefloat{figure}
\begin{center}
\begin{figure}[H] 
\epsfig{figure=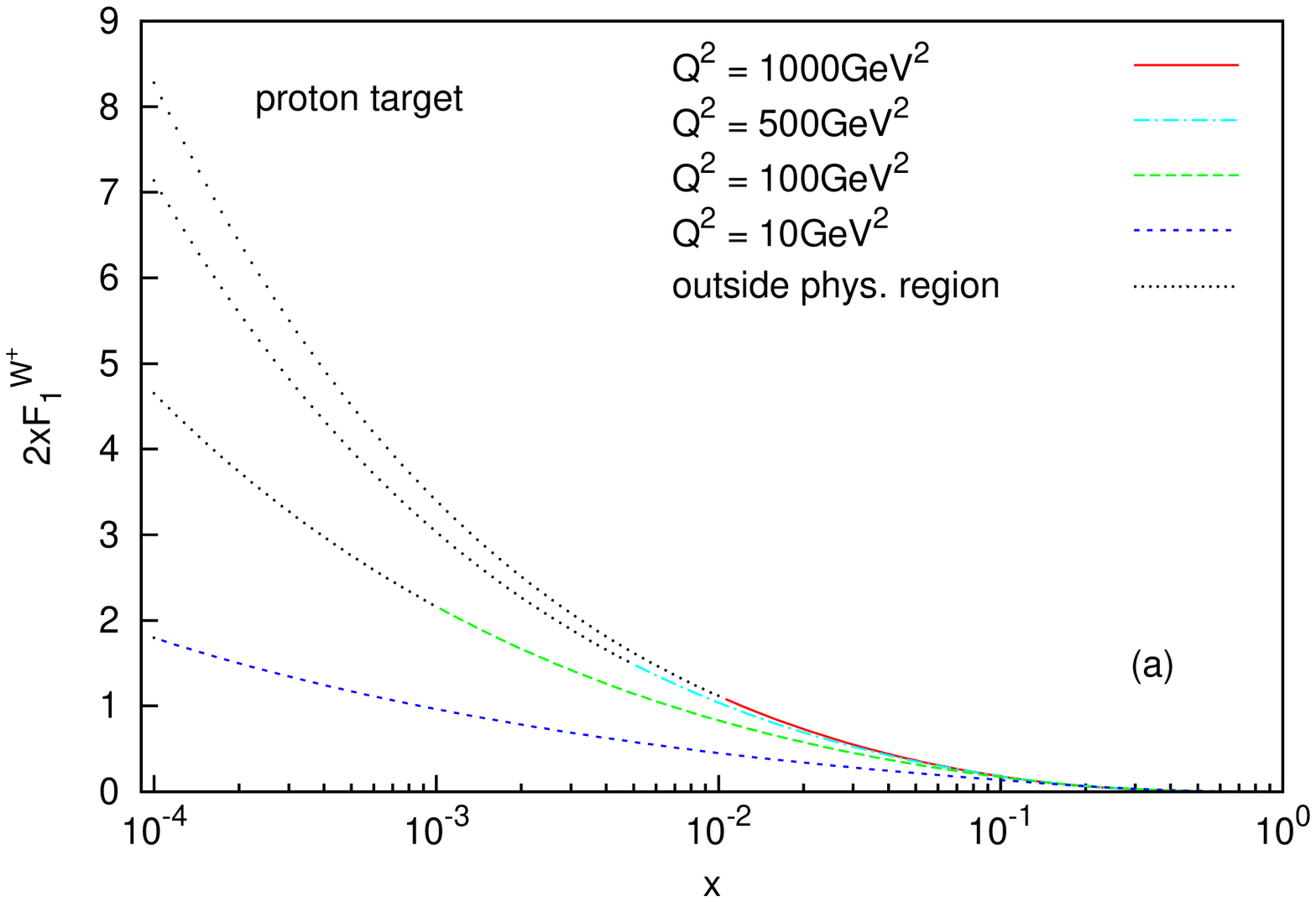,angle=-90,width=0.43\linewidth} \hspace*{2mm}
\epsfig{figure=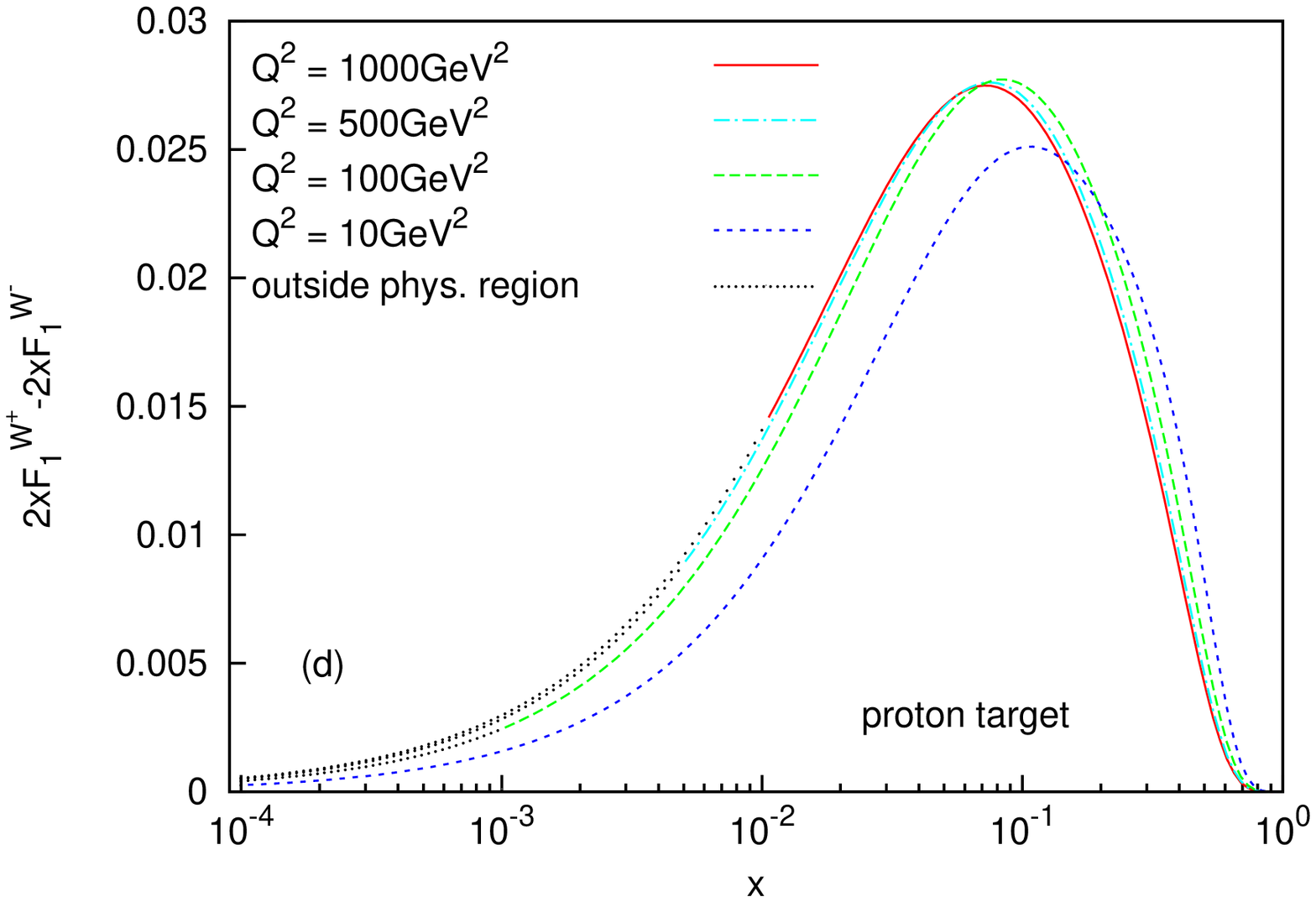,angle=-90,width=0.43\linewidth} 
\\
\epsfig{figure=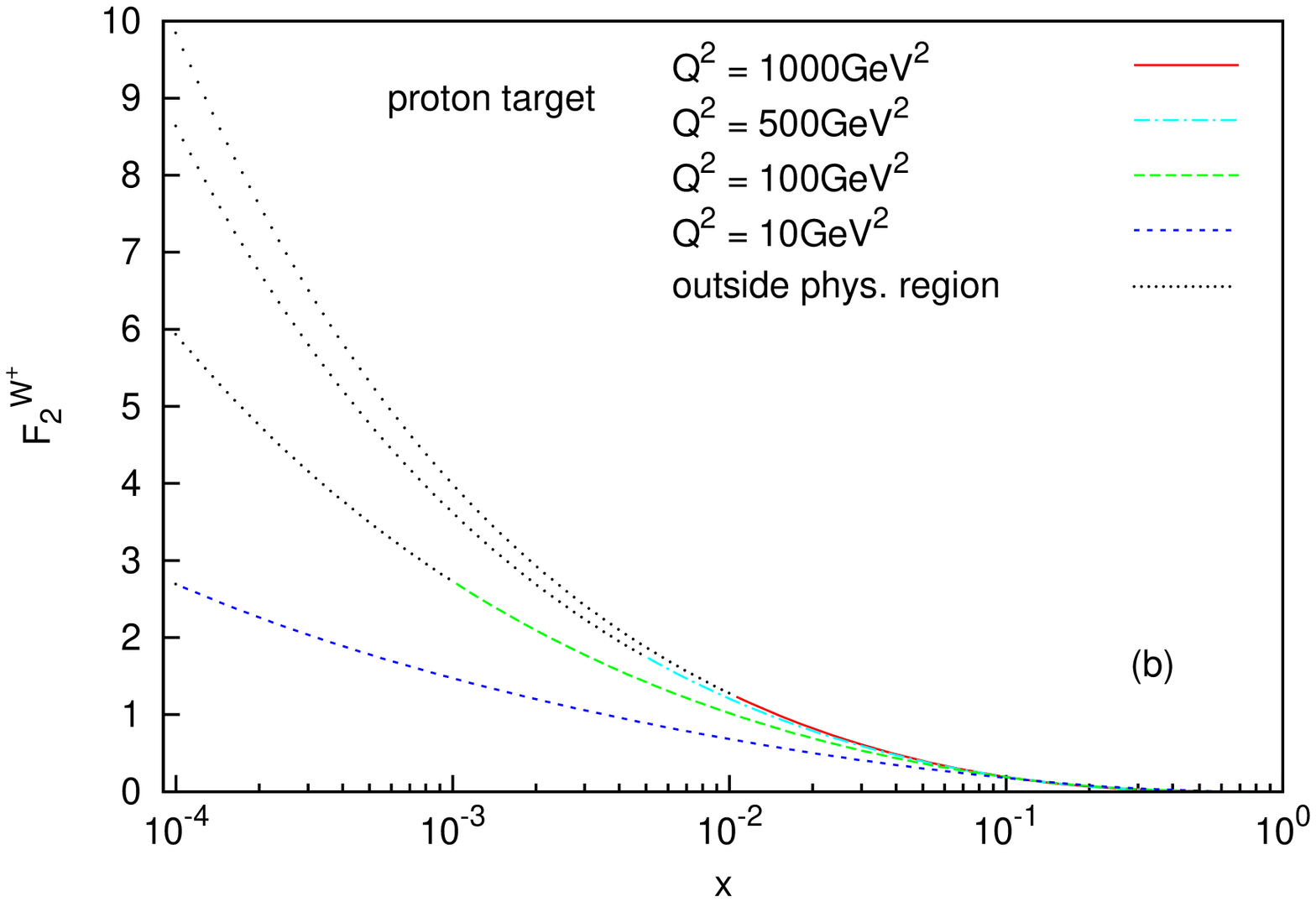,angle=-90,width=0.43\linewidth} \hspace*{2mm}
\epsfig{figure=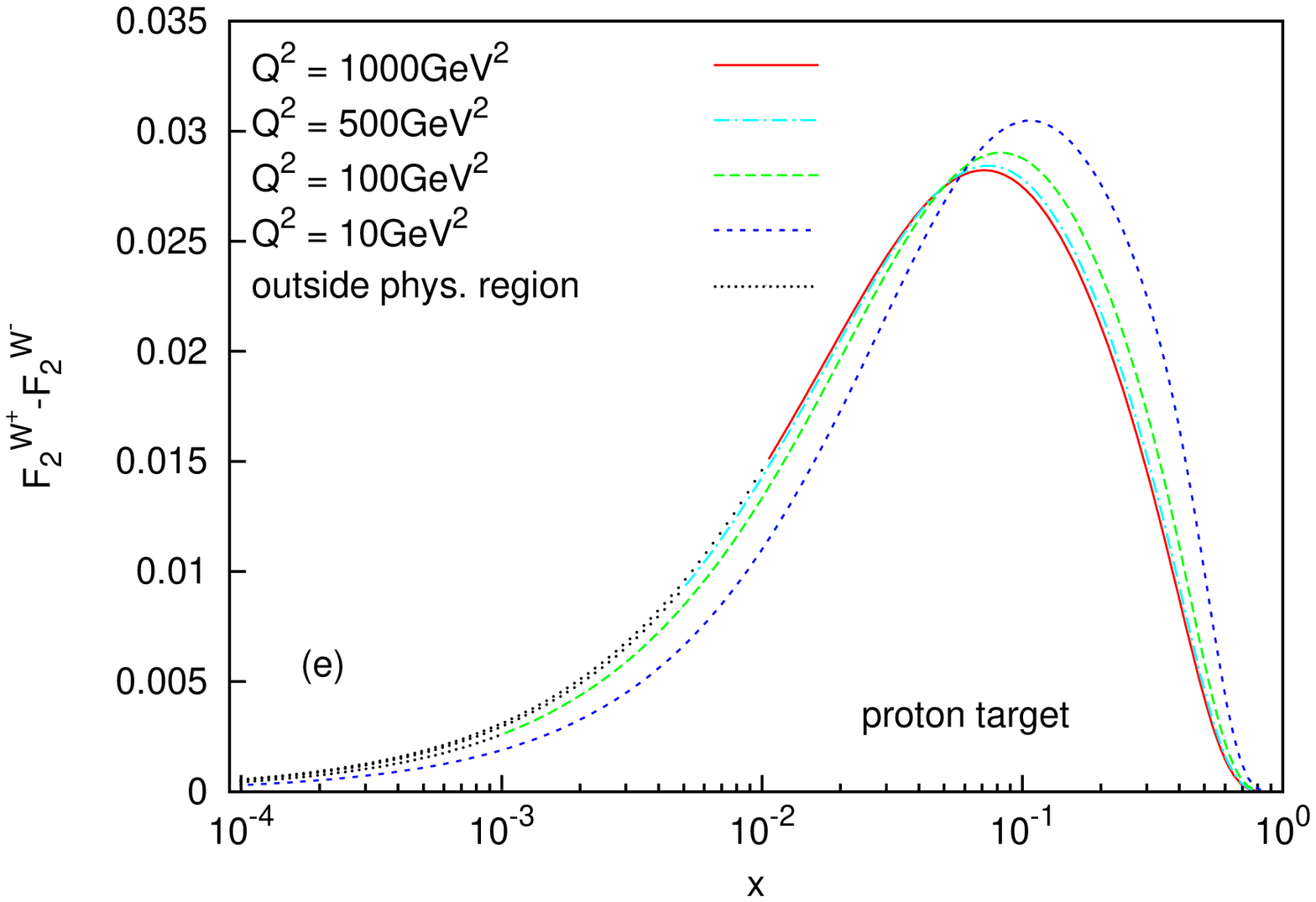,angle=-90,width=0.43\linewidth} 
\\
\epsfig{figure=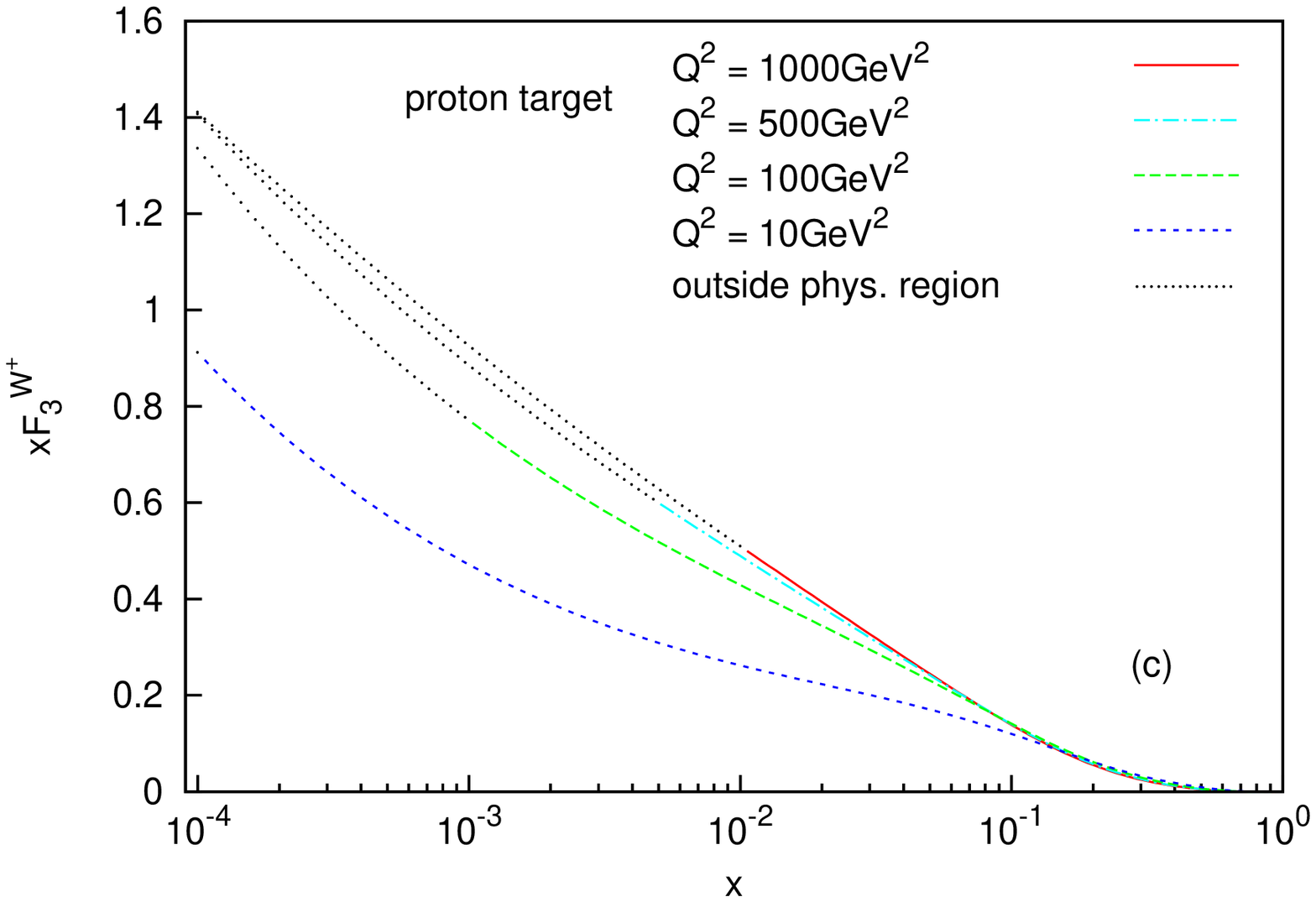,angle=-90,width=0.43\linewidth} \hspace*{2mm}
\epsfig{figure=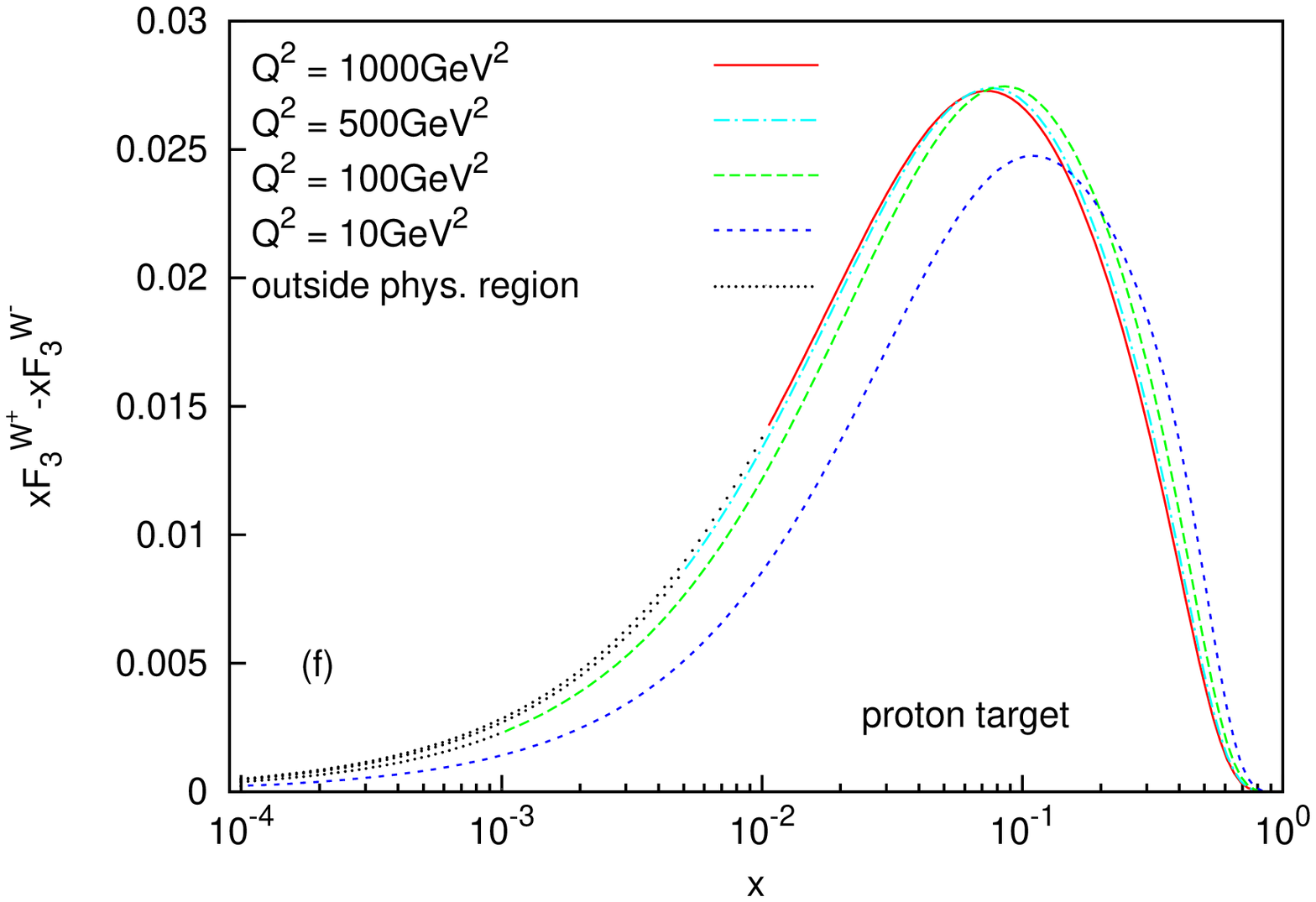,angle=-90,width=0.43\linewidth} 
\caption[]{
\label{FIG:WZ}
\sf
The charged current structure functions for charm production for different 
values of $Q^2$, using the ABKM09 parton parameterizations \cite{ABKM09}. We 
indicated the kinematic range being accessible at HERA. (a)--(c)~: the structure 
functions for $W^+$ exchange. (d)--(f) difference between the structure functions
for $W^+$ and $W^-$ exchange.}
\end{figure}
\end{center}

\section{Numerical Results}

\vspace{1mm}
\noindent
We compare the Mellin space representation given in Section~2 with the representation 
in $x$-space of Ref.~\cite{DORTM1} using the reference distribution
\begin{eqnarray}
   xf(x) = x^{-0.1} (1-x)^5
\end{eqnarray}
for both the quark and gluon densities
and determine the relative accuracies {$|\delta \mathbb{F}_i|/\mathbb{F}_i$} 
for different values of $Q^2$ in the massive Wilson coefficients choosing 
{$m_c = 1.5$}~GeV and the corresponding values of $\alpha_s^{\rm NLO}(Q^2)$ 
\cite{ABKM09}. 

\restylefloat{figure}
\begin{center}
\begin{figure}[t] 
\begin{center}
\epsfig{figure=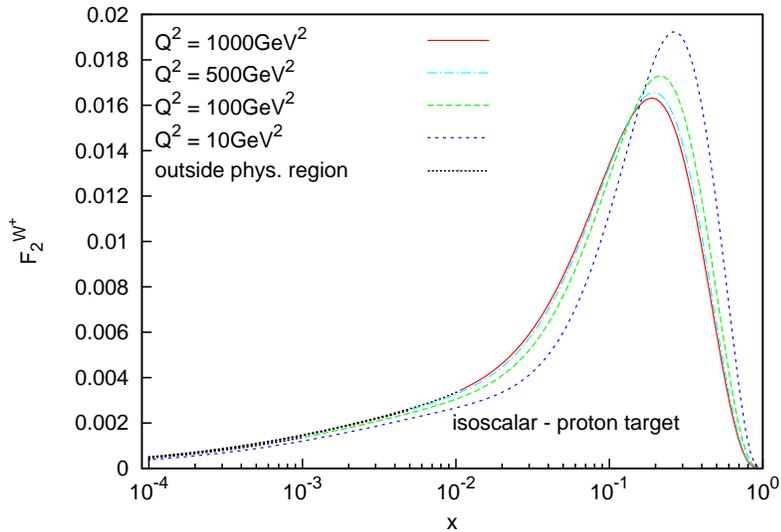,angle=-90,width=0.60\linewidth} 
\end{center}
\caption[]{
\label{FIG:WZ2}
\sf
The difference of the structure function $\mathbb{F}_2$ for $W^+$ exchange for an 
isoscalar and a proton target using the ABKM09 parton distribution functions 
\cite{ABKM09} in depending of $x$ and $Q^2$.
}
\end{figure}
\end{center}

If one employs the {\tt MINIMAX} representation 
relative accuracies of better than $3 \times 10^{-6}$ are reached  
below $x = 0.5$. For $x > 0.5$ the relative accuracy becomes worse. In this 
region, however, the charm contribution is very much suppressed, as shown in 
Figures~2 and 3. Using the representations (\ref{eq:F1U}, \ref{eq:F2U}) the 
relative accuracy is improved and amounts to $5 \times 10^{-8}$ at $x = 10^{-4}$ or 
better growing to $\sim 10^{-6}$ for $x \sim 0.4$, cf. Figure~1,a--c. Beyond this value the 
relative accuracy becomes worse, but also the sea quark distributions are very 
small in this region. For comparison we note that in
\cite{Ball:2011mu} accuracies of 0.015 to 0.002 were obtained.

In Figure~2 we show the charged current heavy flavor structure functions 
$2x\mathbb{F}_1, \mathbb{F}_1$ and $x\mathbb{F}_3$ for 
$ep$-scattering and the kinematics at HERA for $W^+$ exchange and the difference 
of the structure functions for $W^+$ and $W^-$ scattering using 
$m_c = 1.5$ GeV, referring to the ABKM09 NLO 
distributions in the fixed flavor scheme \cite{ABKM09}. Results are shown 
for $Q^2 = 10, 100, 500$ and $1000$ GeV$^2$,
($\alpha_s^{\rm NLO} = 0.2399, 0.1666, 0.1379, 0.1283$).
The dominant contributions are those due to $W^{\pm}$--gluon fusion, 
which is the same for $W^+$ and $W^-$ exchange. All structure functions rise towards 
small values of $x$. The difference between the $W^+$ and $W^-$-exchange structure 
functions, on the other hand, receives its main contributions in the valence region 
with smaller scaling violations as in case of $W^{\pm}$--gluon fusion.  

In Figure~3 the difference of the structure function $F_2^{W^+}$ for an isoscalar 
and proton target is shown. This also is a valence-like function. With the present 
precise Mellin-space implementations at hand one may perform QCD fits including the 
charged current heavy flavor contributions in an efficient way.  
\section{Parameterizations in Mellin Space}

\vspace*{1mm}
\noindent
Parameterization of the non-perturbative input distribution of the parton distribution 
functions at a starting scale $\mu^2_0$ containing sufficient flexibility is an 
important condition for the QCD analyses of deep-inelastic and hard scattering data. In the 
literature different parameterizations of distributions are used. A wide-spread shape has 
the form  
\begin{eqnarray}
f_i(x,\mu_0^2) = A x^\alpha (1-x)^\beta P(x),~~P(x) = \sum_{k=1}^K \gamma_k 
x^{\delta_k}~.
\end{eqnarray}
The corresponding Mellin transform reads
\begin{eqnarray}
\label{eq:PAR2}
f_i(N,\mu_0^2) = A \sum_{k=1}^K \gamma_k B(N+\alpha+\delta_k, \beta+1)~, 
\end{eqnarray}
with parameters such that $N+\alpha+\delta_k, \beta+1 \neq -l,~~ \forall l \in 
\mathbb{N}$ and
$B(x,y) = \Gamma(x) \Gamma(y)/\Gamma(x+y)$. Eq.~(\ref{eq:PAR2}) is analytic under this 
condition.

One may adopt the attitude to represent the non-perturbative momentum distributions 
of 
the partons, which are $L^2[0,1]$ measurable functions, in terms of orthogonal 
polynomials, see \cite{ORTHPOL,FP}. If the argument of these polynomials is $x$ or 
a real power of $x$ one may refer to structures as (\ref{eq:PAR2}), provided the 
Mellin 
transforms exist. The fastest convergence is obtained choosing
Laguerre polynomials \cite{LAGUERRE} of the argument $\ln(1/x)$, \cite{FP}. 
\begin{eqnarray}
\label{eq:PAR2a}
f_i(x,\mu_0^2) = \sum_{k=0}^\infty c_k 
L_k\left(\ln\left(\frac{1}{x}\right)\right)~,
\end{eqnarray}
with
\begin{eqnarray}
\label{eq:PAR2a1}
L_k(z) = \sum_{l=0}^k \binom{k}{l} (-1)^l 
\frac{z^l}{l!}~.
\end{eqnarray}
The Mellin transform reads 
\begin{eqnarray}
\label{eq:PAR2b}
f_i(N,\mu_0^2) = \frac{1}{N} \sum_{k=0}^\infty c_k \left(\frac{N-1}{N}\right)^k~. 
\end{eqnarray}

However, also other parameterizations are used in $x$-space analyses. One of them  
\cite{AMP,ABKM09} refers to the following shapes
\begin{eqnarray}
f_i(x,\mu_0^2) = A x^\alpha (1-x)^\beta x^{a x(1 + b x)},
\end{eqnarray}
partly with $b = 0$. Let us first consider the simplified case $b = \beta = 0$ 
as an illustration. Here, one obtains
\begin{eqnarray}
f_i(N,\mu_0^2) = A \sum_{k=0}^\infty \frac{(-a)^k}{(N+\alpha+k)^{k+1}}~,
\end{eqnarray}
which is easily recognized as a quickly convergent series in $k$. In the general case we 
obtain
\begin{eqnarray}
\label{eq:REP3}
f_i(N,\mu_0^2) = A \sum_{k=0}^\infty \frac{a^k}{k!} \sum_{l=0}^k 
\binom{k}{l} b^l
\frac{\partial^k}{\partial N^k}
B(N + \alpha + k + l, \beta + 1).
\end{eqnarray}
The derivatives of the Beta-function in (\ref{eq:REP3}) can be calculated in a 
recursive way, see \cite{HSUM2}. Let
\begin{eqnarray}
\label{eq:REP3a}
\frac{1}{k!}
\frac{\partial^k}{\partial N^k} B(N + \alpha, \beta) = B(N + \alpha, \beta)~\Delta_k,~~
\Delta_0 =1~. 
\end{eqnarray}
Then one obtains
\begin{eqnarray}
\Delta_k = \frac{1}{k} \sum_{l=1}^k \Delta_{k-l} f_l~,
\end{eqnarray}
with 
\begin{eqnarray}
f_l = \frac{1}{(l-1)!} \left[\psi^{(l-1)}(N + \alpha) - \psi^{(l-1)}(N + \alpha + 
\beta)\right]~.
\end{eqnarray}

Yet another parameterization was used in \cite{CTEQ},
\begin{eqnarray}
\label{eq:DIS1}
f_i(x,Q^2) =  A x^\alpha (1-x)^\beta e^{\gamma x} (1 + \delta x)^\eta.  
\end{eqnarray}
We first consider the case $\delta = 0$. The Mellin transform of (\ref{eq:DIS1}) is 
then given by
\begin{eqnarray}
\label{eq:DIS2}
f_i(N,Q^2) &=&  
A~B(N+\alpha,\beta+1)~_1F_1(\alpha+N, \beta+\alpha+N+1; \gamma) \nonumber\\
&=&
A~B(N+\alpha,\beta+1)~e^\gamma~_1F_1(\beta+1,N+\alpha+\beta+1; -\gamma)~,
\end{eqnarray}
where $_1F_1(\alpha, \beta; z)$ denotes the confluent hypergeometric function
\cite{KUMMER}. It obeys the recursion relation \cite{NIST}
\begin{eqnarray}
\label{eq:DIS2a}
 _1F_1(\alpha, \beta+1; z) = - \frac{1}{z(\beta-\alpha)} \left[ \beta(1-\beta-z) 
_1F_1(\alpha, \beta; z) 
+ \beta (\beta-1) _1F_1(\alpha, \beta-1; z)\right]~.
\end{eqnarray}
For large values of $N$ the asymptotic relation \cite{NIST,TEMME} 
\begin{eqnarray}
\label{eq:DIS3}
 _1F_1(\beta+1,N+\alpha+\beta+1; -\gamma) &=& \sum_{s=0}^{n-1} 
\frac{\Gamma(\beta+1+s) \Gamma(N+\alpha+\beta+1)}
     {\Gamma(\beta+1) \Gamma(N+\alpha+\beta+s+1)} \frac{(-\gamma)^s}{s!}
\nonumber\\ &&
+ O\left(|N+\alpha+\beta+1|^{-n}\right)
\end{eqnarray}
holds. Eqs.~(\ref{eq:DIS2a},\ref{eq:DIS3}) may thus be used to compute the analytic 
continuation of (\ref{eq:DIS2}).

For the additional factor
\begin{eqnarray}
(1 + \delta x)^\eta
\end{eqnarray}
one may extract an integer power from $\eta$ such that  the 
remainder,$\bar{\eta}$, obeys 
$\bar{\eta} \in [-1/2, 1/2]$. The integer contribution $\tilde{\eta} = \eta - 
\bar{\eta}$ is modifying the above representations but leads to the same structure.
The remaining factor possesses the well converging representation 
\begin{eqnarray}
(1 + \delta x)^{\bar{\eta}}  = \sum_{l=1}^\infty \binom{\bar{\eta}}{l} 
\delta^l~x^l~,
\end{eqnarray}
and
leads to a corresponding generalization of relations~(\ref{eq:DIS2}, \ref{eq:DIS3}).
\section{Conclusions}

\vspace*{1mm} 
\noindent 
A fast and precise Mellin-space implementation of the 
$O(\alpha_s)$ massive charged current Wilson coefficients for deep-inelastic 
scattering is provided. This process is of relevance for the understanding of 
di-muon production in deep-inelastic neutrino-nucleon scattering and for the 
deep-inelastic charged current data measured at HERA in particular w.r.t. the 
extraction of the unpolarized strange-quark distribution. To represent the Wilson 
coefficients we used both the {\tt MINIMAX}-method and completely analytic 
representations and exploited recurrences in the complex-valued Mellin variable $N$ 
for $N \rightarrow N+1$. In wide kinematic ranges relative accuracies of $10^{-7}$ 
and better are obtained for the deep-inelastic structure functions. Errors in the 
literature have been corrected. We present phenomenological applications for the 
structure functions $\mathbb{F}_i(x,Q^2),~~i = 1,2,3$ both for $W^+$ and $W^-$ 
exchange. Furthermore, we provided Mellin-space representations for a wide range of 
representations used in different $x$-space analyses. This allows both for more 
flexible choices of the parton distributions at the input scale using Mellin-based 
codes in data analyses and for comparisons with fits given in the literature. 
The corresponding code is based on {\tt ANCONT} \cite{ANCONT} and available on 
request from {\tt Johannes.Bluemlein@desy.de}.

\vspace*{4mm}
\noindent
{\bf Acknowledgment.}~
{We would like to thank S. Alekhin for discussions and  N. Temme for useful remarks.
This paper has been supported in part by DFG Sonderforschungsbereich Transregio 9,
Computergest\"utzte Theoretische Teilchenphysik and EU Network {\sf LHCPHENOnet}
PITN-GA-2010-264564.}


\end{document}